\documentclass{osa-article}
\newcommand {\ifHLcorrections} {\iffalse} 

\usepackage{graphicx}
\makeatletter
\let\saved@includegraphics\includegraphics
\AtBeginDocument{\let\includegraphics\saved@includegraphics}
\renewenvironment*{figure}{\@float{figure}}{\end@float}
\makeatother
\usepackage{mathtools}
\usepackage{siunitx}
\usepackage{color}
\usepackage[normalem]{ulem}

\articletype{Research Article}

\begin{document}

\title{High Quantum Efficiency Photoconductive Heaters Enable Control of Large-Scale Silicon Photonic Ring Resonator Circuits}

\author{Hasitha Jayatilleka\authormark{1,2,3,*}, Hossam Shoman\authormark{1,3,*}, Lukas Chrostowski\authormark{1}, and Sudip Shekhar\authormark{1}}

\address{\authormark{1} Department of Electrical and Computer Engineering, University of British Columbia, 2332 Main Mall, Vancouver, British Columbia V6T 1Z4, Canada.\\
\authormark{2} Currently with Intel Corporation, 2200 Mission College Blvd, Santa Clara, California, 95054, USA. \\ \authormark{3} Contributed equally to this work.}

\email{\authormark{*}hasitha@ece.ubc.ca, hoshoman@ece.ubc.ca} 


\begin{abstract}
A multitude of large-scale silicon photonic systems based on ring resonators have been envisioned for applications ranging from biomedical sensing to quantum computing and machine learning. Yet, due to the lack of a scalable solution for controlling ring resonators, practical demonstrations have been limited to systems with only a few rings. Here, we demonstrate that large systems can be controlled only by using doped waveguide elements inside their rings whilst preserving their area and cost. We measure the large photoconductive changes \ifHLcorrections{\sout{\color{red}{(quantum efficiencies $\sim$10)}}}\fi of the waveguides for monitoring rings' resonance conditions across high-dynamic ranges and leverage their thermo-optic effects for tuning. This allows us to control ring resonators without requiring additional components, complex tuning algorithms, or additional electrical I/Os. We demonstrate automatic resonance alignment of $31$ rings of a $16\times16$ switch and of a $14$-ring coupled resonator optical waveguide (CROW), making them the largest, yet most compact, automatically controlled silicon ring resonator circuits to date.
\end{abstract}

\section{Introduction}

The miniature size of silicon ring resonators make them attractive candidates for large-scale photonic systems as they can be densely integrated on-chip for lowering size, power-consumption, and cost\cite{Batten2009-ke,Cheng2018-ct, Dong2018-it}. As a result, numerous solutions based on ring resonators have been proposed for applications in communications systems\cite{Atabaki2018-dv,Dong2015-se,Cheng2018-ct,nikolova2017modular}, signal processing\cite{Batten2009-ke,Khilo2012-rk}, quantum computing\cite{Kumar2014-yi}, sensing\cite{Wang2014-tl}, and machine learning\cite{Tait2017-so}.
A key requirement for the practical use of these systems is the ability to precisely control the resonance conditions of their ring resonators, which allows to 1) correct for fabrication errors, 2) adapt the system in real-time to account for temperature variations or laser wavelength fluctuations, and 3) reprogram the system altogether for implementing various transfer functions and different functionalities. Such control can be enabled by utilizing feedback loops to monitor and tune/track the resonance conditions of the rings until the desired conditions are met.  Prior-art for controlling ring resonators have depended on numerous photodetectors (PDs) for monitoring the rings' resonance conditions and on separate thermo/electro-optic phase shifters for tuning the rings 
\cite{Padmaraju2014-fq,Zhang2014-hf,Li2015-mz,Mak2015-wj,Grillanda2014-nf,Wang2017-uv}. However, these extra components often need additional processing steps (e.g., Ge depositions\cite{Mak2015-wj, Poulton} or Si+ implantations\cite{Geis2009-xt} for PD), increase the number of electrical inputs/outputs to the system, and occupy significantly large on-chip real estate\cite{Padmaraju2014-fq,Grillanda2014-nf,Mak2015-wj,Li2015-mz}. Therefore, a low-cost single element which can be placed inside the resonator for monitoring and tuning its resonance would be a true enabler for controlling large-scale systems.

In this article, we discuss the physics, report on the photodetection quantum efficiencies (QEs), and demonstrate the capabilities of such a control element in tuning large-scale silicon ring resonator systems.  
The control element we demonstrate is based on a doped silicon nanowire waveguide, which is ubiquitously found across many industrial silicon photonics platforms without additional process modifications. 
We combine the doped waveguides' photoconductive effects  together with their thermo-optic tuning capabilities for monitoring and tracking the ring resonators' resonance conditions, respectively. Previously, we used similar doped waveguides for controlling single and coupled ring resonators \cite{Jayatilleka2015-hw, Jayatilleka2018-ed}. Yet, the scalability of such solutions towards large systems was unclear due to the unkown detection capabilities such as the QEs and dynamic ranges. Here, we demonstrate capabilities of our photoconductive heaters, to the best of our knowledge, by automatically aligning the resonances of the largest number of rings on a silicon chip, i.e., 31 rings along the longest path of a $16\times16$  switch and a 14-ring CROW, respectively. While these systems show large insertion loss variations, the precise resonance detection is enabled by the large {\ifHLcorrections{\color{red}\fi 43 dB } dynamic range   permitted by the record high photoconductive QE of the doped waveguides. Almost all ring resonator-based systems are formed by using two types of circuits as building-blocks: 1) rings inter connected through bus waveguides\cite{Dong2015-se, Tait2017-so, Batten2009-ke, Atabaki2018-dv,  nikolova2017modular, khope2017chip} and/or 2) rings coupled to each other (i.e., CROWs)\cite{Kumar2014-yi,Wang2014-tl,Aguiar2018-zy,Jayatilleka2018-ed,Morichetti2012-hz}. In this article, by demonstrating the automatic control of the largest number of resonators in each type of circuit, we show how {\ifHLcorrections{\color{red}\fi such photoconductive heaters} can be readily deployed to control a majority of the ring resonator systems proposed before.

{\ifHLcorrections{\color{red}\fi
The rest of this paper is organized as follows. In section 2, we discuss the photoconductive and thermo-optic behaviors of the doped silicon waveguides demonstrating their high QEs, large dynamic ranges, photodetection bandwidths and thermo-optic tuning bandwidths. We then show how they can be integrated into ring resonators for monitoring and tracking the rings' resonance conditions. In section 3, we  demonstrate tuning the ring resonators along the longest path of a $16\times16$  switch, routing light through 31 ring resonators connected through bus waveguides and 30 waveguide crossings. In section 4, we demonstrate the automatic resonance alignment of a CROW with 14 ring resonators.}

\section{Silicon waveguide photoconductive heaters} 

\subsection{Photoconductive and thermo-optic behaviors of doped waveguides}

\begin{figure}[b]
	\centering
	\includegraphics[width=\textwidth]{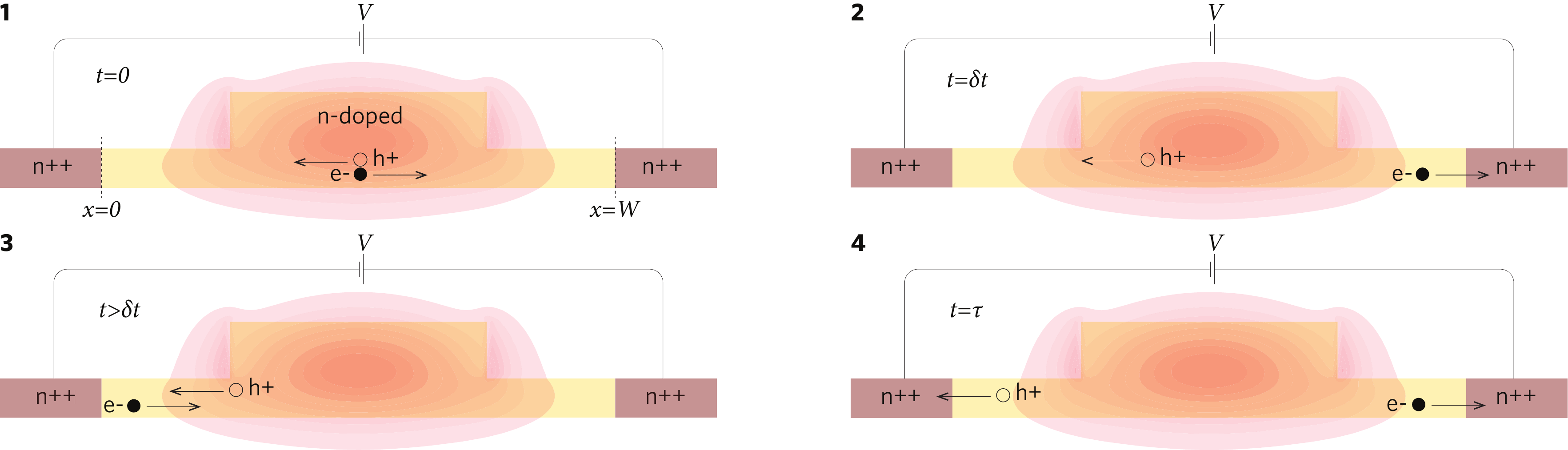}
	\caption{{Photoconductivity of a silicon nanowire waveguide.} Cross sections of a silicon waveguide illustrating the transport of a photogenerated electron and hole towards the terminals at different velocities. The additional electrons injected from the negative terminal in order to maintain the  charge neutrality of the semiconductor results in a large net gain in the QE.  The centre of the waveguide is lightly doped ($5\times10^{17}$ cm$^{-3}$), whereas the sides are highly doped to form ohmic contacts. The overlap of the simulated TE optical mode at \SI{1.55}{\micro\meter} with the waveguide is also shown.}
	\label{fig1}
\end{figure}

Fig.~\ref{fig1}a illustrates the photoconductive mechanism of the doped silicon nanowire rib waveguides used in this work. In contrast to the previous reports\cite{Zhou2014-yj,Geis2009-xt,Baehr-Jones2008-ez,Li2015-mz}, the key enabler in our design is the lightly n-doped ($5\times10^{17}$ cm$^{-3}$) waveguide core. In previous reports on silicon waveguide photodetectors, the waveguide was left undoped\cite{Baehr-Jones2008-ez,Zhou2014-yj} or was implanted with Si+ ions\cite{Li2015-mz,Geis2009-xt}, which either reduced the measurable photocurrent or increased the loss of the waveguide, respectively. Furthermore, a vast majority of previous reports relied on pn-junctions\cite{Geis2009-xt, Li2015-mz, zhang2018enhancing} for measuring photocurrents which limited their use as phase tuners due to sub-nm tuning ranges.
In our design, we only use n-type doping. While the doping in the waveguide core is low enough to allow for low-loss (doping loss = 5 dB/cm) propagation, it is sufficient enough to 1) lower the electrical resistance across the waveguide enabling the device to function as a thermo-optic tuner over appreciable wavelength ranges with low voltages compatible with CMOS circuitry and to 2) increase the measurable photocurrent to micro-Amperes (for micro-Watt input optical powers) allowing the device to also function as a precise power monitor inside an optical circuit. We attribute the generation of the initial electron-hole pairs (EHPs) inside the waveguide to absorption in the small number of defect (due to doping) and surface states. As a result of the mobility difference between electrons and holes, electrons reach the positive terminal before holes reach the negative terminal (See Fig. 1). Additional electrons are injected into the device from the negative terminal to maintain the semiconductor's charge neutrality and, hence, multiple electrons traverse across the device in the time it takes a photo-generated hole to reach the negative terminal. As a result, the number of collected electrons across the terminals far exceed the number of photo-generated EHPs, resulting in a high photoconductive gain in the QE yielding appreciable photocurrents with low-loss.

\begin{figure}[h]
	\centering
	\includegraphics[width=\textwidth]{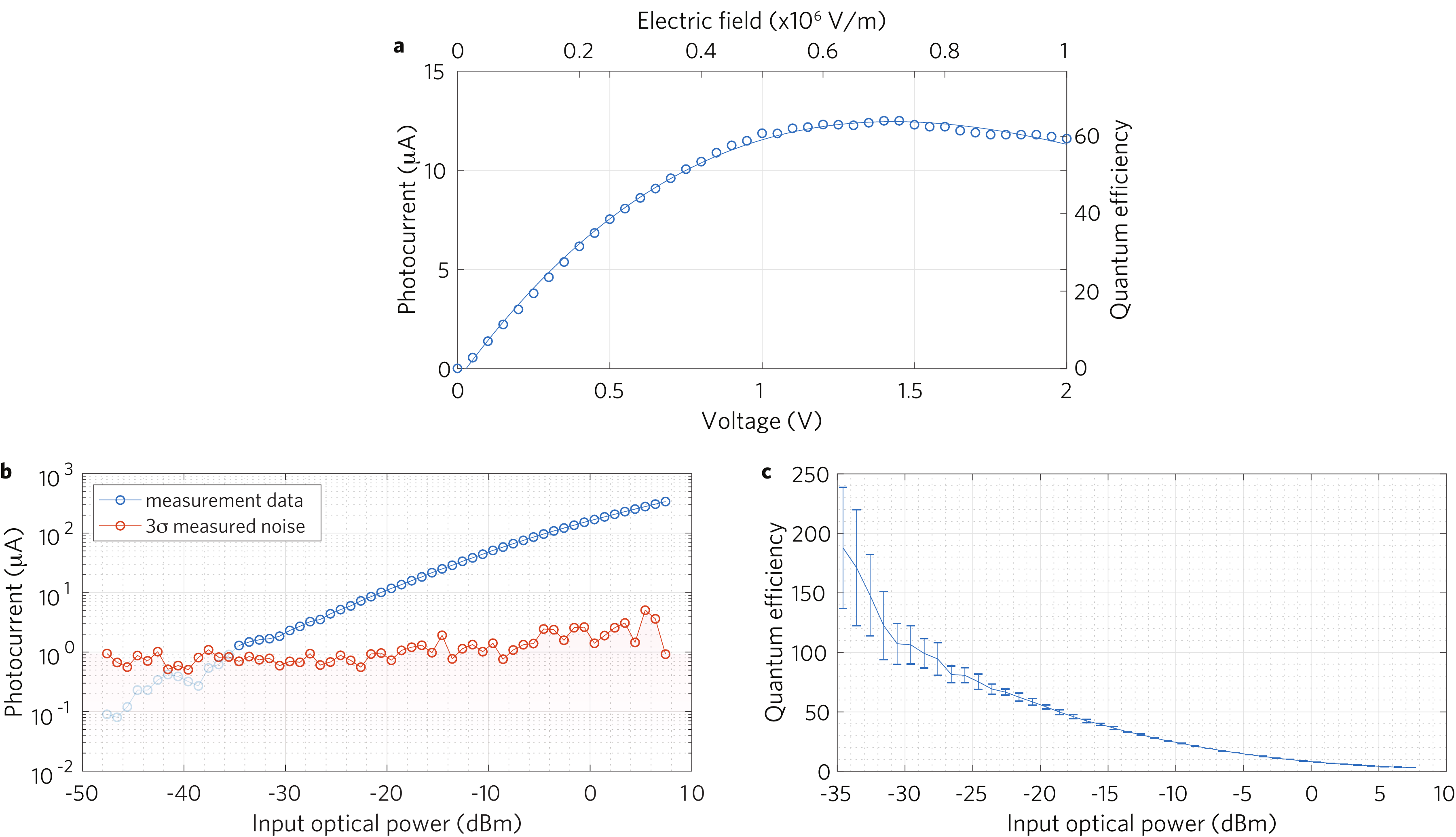}
	\caption{Photoconductivity response of a silicon nanowire waveguide. Photocurrent and QE measured from a \SI{100}{\micro\meter} long photoconductive waveguide as functions of the (a) bias voltage (at an input power of \SI{10}{\micro\watt}). The large gain in QE reduces beyond $1.4$ V due to the electrons and holes reaching their respective saturation velocities. The solid line is a polynomial fit to the measurement.  Measured (b) Photocurrent and (c) QE as functions of the  input optical power (at a bias voltage of $1$ V). In (b), $\pm3\sigma$ variation of the the $I_\text{PD}$ measurement at each input optical power is shown. In (c), the QE reduces with the input optical power due to saturation of the defect and surface states. The error bars shown in (c) correspond to the error in QE due to the $\pm3\sigma$ current measurement noise.}
\label{figirphs2}
\end{figure}

\begin{figure}[h]
	\centering
	\includegraphics[width=\textwidth]{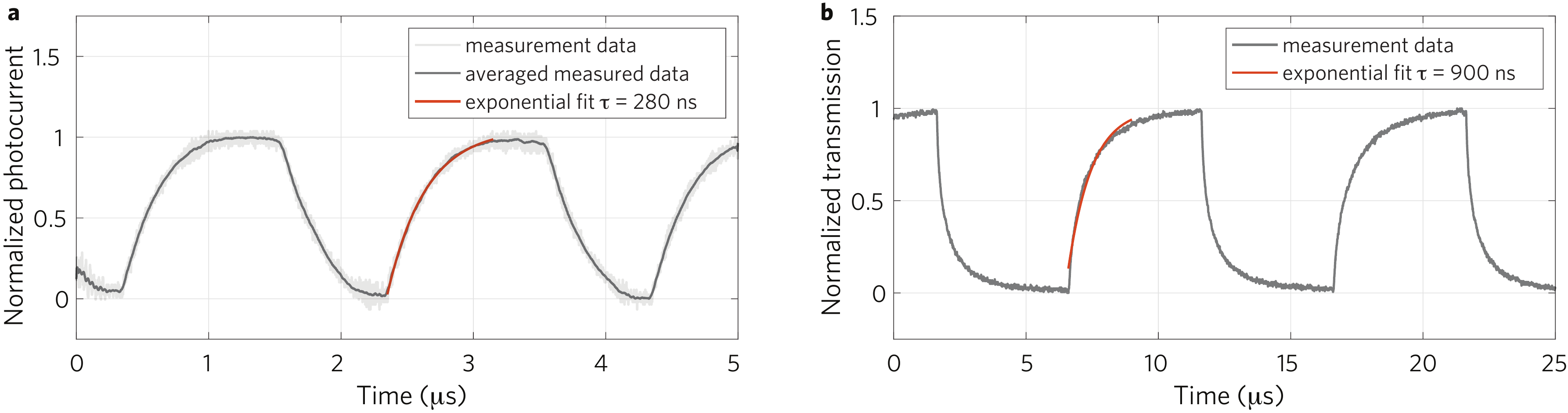}
	\caption{{Time-domain photodetection and thermo-optic tuning responses of a photoconductive heater. (a)} Photodetection response of the device heater when the input light is modulated at $500$ kHz. The rise-time of the response is \SI{0.28}{\micro\second}, corresponding to a photodetection bandwidth of \SI{570}{kHz}. (b) Thermo-optic response of the photoconductive heater measured by modulating a device integrated into Mach-Zehnder interferometer with a $100$ kHz square wave signal. The rise-time of the response is \SI{0.9}{\micro\second}, corresponding to a $3$-dB thermo-optic tuning bandwidth of \SI{175}{kHz}.}
	\label{figirphs3}
\end{figure}

Fig.~\ref{figirphs2}a shows the measured photocurrent ($I_\text{PD}$) and QE for a \SI{100}{\micro\meter} long n-type photoconductive heater at an input power and wavelength of \SI{10}{\micro\watt} and \SI{1.55}{\micro\meter}, respectively, as a function of the applied {\ifHLcorrections{\color{red}\fi heater bias ($V_\text{heater}$)}. First, the dark-current ($I_\text{heater}$) was measured with the input laser turned off (see \textcolor{urlblue} {Supplementary Fig. S2c}). $I_\text{PD}$ was then obtained by subtracting the calibrated $I_\text{heater}$ from the total measured current when the input laser is on. \ifHLcorrections{\color{red}\sout{The QE was calculated assuming that all of the lost photons due to doping (5 dB/cm) and scattering (1.9 dB/cm) generated EHPs.}}\fi{\ifHLcorrections{\color{red}\fi When calculating QE, we estimated the total insertion loss of the device to be 0.069 dB arising from a  doping loss of $5$ dB/cm and a scattering loss of $1.9$ dB/cm.} Therefore, the QE reported here represents the smallest possible QE (see \textcolor{urlblue}{Supplementary section 2}). In Fig.~\ref{figirphs2}a, the QE peaks to $64$ at $1.4$ V and starts to diminish for higher bias voltages. This is because the difference between electron and hole mobilities diminish at high fields reducing the photoconductive gain\cite{Kasap2013-sw}. Fig.~\ref{figirphs2}b shows the measured photocurrents and {\ifHLcorrections{\color{red}\fi Fig.~\ref{figirphs2}c shows the corresponding QEs as functions of the optical power entering a similar device at a bias of $1$ V}. The QE reduces for high optical powers due to the saturation of the defect and surface states in the waveguide.  
{\ifHLcorrections{\color{red}\fi Fig.~\ref{figirphs2}b shows detection of optical powers ranging from -35 dBm to 8 dBm across a 43 dB dynamic range. The measurable dynamic range here was limited by the largest input optical power and the smallest measurable $I_\text{PD}$ as allowed by our setup. The smallest measurable $I_\text{PD}$ was limited due to the measurement error of our source-measure unit, where we estimated the standard deviation $\sigma$ value to be \SI{0.15}{\micro A} when measured at a bias of 1 V over an integration time of \SI{416.7}{ms} per measurement. In the experiments described in Sections 3 and 4, we used an integration time of \SI{16.7}{ms} for photocurrent measurement.} The $\sigma$ value for each optical input power was calculated from 50 photocurrent measurements (see \textcolor{urlblue} {Supplementary Fig. S2d} for $\sigma$ calculation of the dark current $I_{heater}$). In Fig.~\ref{figirphs2}b we set the noise floor to $\pm 3 \sigma$ (\SI{0.9}{\micro A}) for reporting photocurrents and QE values.

Photoconductive heater's detection bandwidth depends on the electron and hole transit times across the waveguide. 
Fig.~\ref{figirphs3}a shows the dynamic response of a photoconductive heater measured by modulating the input light with a \SI{500}{kHz} square wave. 
We extracted the rise- and fall-times and the $3$-dB photodetection bandwidth \cite{watts2009adiabatic} of the device to be {\SI{0.28}{\micro\second}  and $570$ kHz, respectively. 
{\ifHLcorrections{\color{red}\fi Fig.~\ref{figirphs3}b shows the dynamic thermo-optic response measured by modulating a similar photoconductive heater embedded in a Mach-Zehnder interferometer with a $100$ kHz square wave voltage signal. The $3$-dB thermo-optic tuning bandwidth was extracted to be \SI{175}{kHz}. 
As the $3$-dB photodetection bandwidth is much greater than the thermo-optic bandwidth, loop bandwidth of a feedback-based ring resonance tracking system using integrated CMOS electronic circuits will likely be limited by the thermo-optic tuning speed.
}

%
%

\subsection{Ring resonator control}

Fig.~\ref{fig2}a shows the integration of a photoconductive heater into a silicon ring resonator of \SI{8}{\micro\meter} radius by n-doping a quarter of the ring's perimeter. 
The ring's intracavity optical power can be measured (via $I_\text{PD}$) and the resonance can be tuned (via $I_\text{heater}$) using this single element. 
The same contact pad is used for measuring $I_\text{PD}$ and supplying the heater current, $I_\text{heater}$, for tuning. Therefore, additional electrical I/Os are not necessary to fully control the ring by using a feedback loop for sensing and tuning its resonance.  Fig.~\ref{fig2}b shows the measured drop-port transmission and photocurrent, $I_\text{PD}$, as a function of the supplied electrical power to the heater. {\ifHLcorrections{\color{red}\fi The electrical power to the heater is controlled by setting the heater voltage $V_\text{heater}$.}
As compared to a straight waveguide, the photocurrent measured from the ring is further amplified due to the energy build-up inside the cavity. 
{\ifHLcorrections{\color{red}\fi Both the drop-port transmission, which is proportional to the ring's intracavity optical power,  and the measured $I_\text{PD}$ changes in sync with each other and they are both maximized at the same electrical power. Hence, the ring can be set to be resonant with the input laser's wavelength by maximizing the photocurrent, $I_\text{PD}$. 
If the resonator's bandwidth is extremely narrow, behavior of photoconductivity  as function of the heater voltage, $V_\text{heater}$,  (see Fig. \ref{figirphs2}a) can introduce a deviation between the $V_\text{heater}$ values that maximize the drop-port response and  $I_\text{PD}$.  In such cases, the measured  $I_\text{PD}$ values can be calibrated to account for the change in photoconductivity  as function of the heater's voltage  (see Fig. \ref{figirphs2}a).
Due to the large optical pass bandwidths (e.g., 3-dB bandwidth of the resonator shown in Fig.~\ref{fig2} is 44 GHz ) of the resonators used in this work, we did not observe any appreciable deviation between $V_\text{heater}$ values that maximized the drop-port response and  $I_\text{PD}$. In addition, $I_\text{PD}$ is also temperature dependent. However, maximum seeking algorithms used in this work can overcome such a level shift in $I_\text{PD}$ caused by a temperature drift that is much slower than the tracking speed. In a previous work, we demonstrated tracking the response of a 4-ring CROW across a \SI{65}{\celsius} temperature variation \cite{Jayatilleka2018-ed }. }

%
%
The ring resonator shown in  Fig.~\ref{fig2}a requires $48$ mW for tuning across the $12.1$ nm FSR. 
In a commercial application, this tuning power can be further reduced by about an order of magnitude by selective silicon substrate removal techniques\cite{Dong2010-yq}. While the intrinsic $\mathcal{Q}$-factor of this ring is limited to about $1.4\times10^5$ due to the light doping, the doping concentration or the length of the doped portion in the ring can be traded-off in applications requiring higher $\mathcal{Q}$-factors. {\ifHLcorrections{\color{red}\fi The length of photoconductive heater in the ring is chosen to be \SI{12.55}{\micro \meter} to allow for thermo-optic tuning across the entire FSR within the supply voltage range. The lower limit for the ring resonator's radius for avoiding excessive bending losses for the waveguide geometry (Fig. \ref{fig1}) is about \SI{8}{\micro \metre} \cite{jayatilleka2016}.}

\begin{figure}[t]
	\centering
	\includegraphics[width=\textwidth]{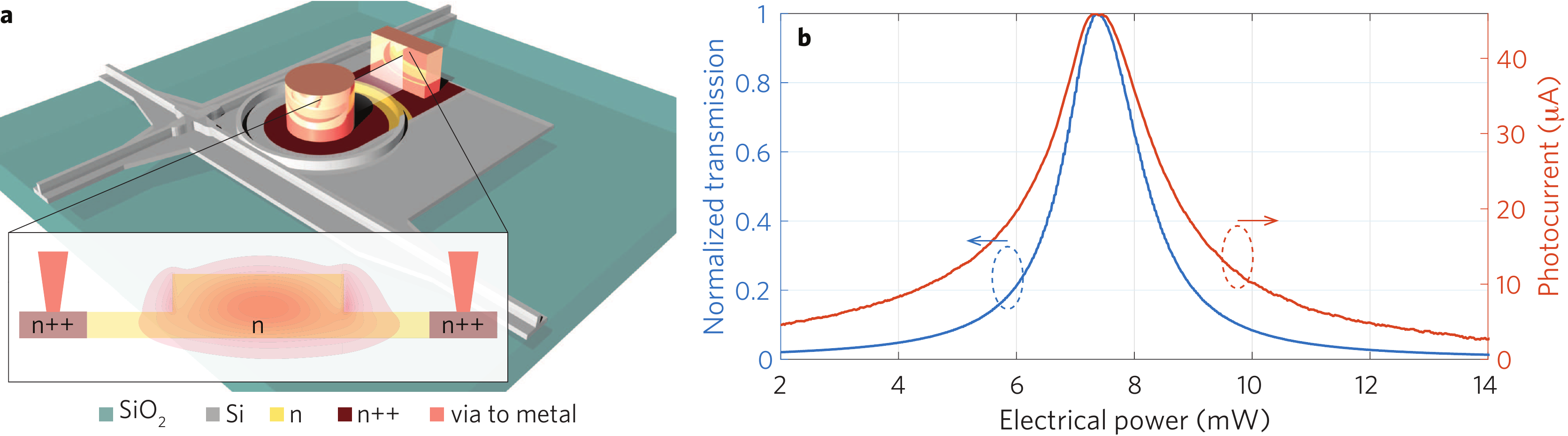} 
	\caption{Integrated photoconductive heaters in a silicon ring resonator. (a) Illustration of a photoconductive heater integrated into the ring resonator with \SI{8}{\micro\meter} radius (drawn to scale). The top SiO$_2$ cladding is not shown. The inset shows an overlap of the simulated TE optical mode at \SI{1.55}{\micro\meter} in the bent waveguide. (b) Measured drop-port transmission (left axis) and photocurrent from the photoconductive heater (right axis) of the ring as a function of the supplied electrical power to the heater. The ring can be set to be resonant with the input light by maximizing the photocurrent.}
	\label{fig2}
\end{figure}

\section{Tuning of switch matrix} 
Ring resonator-based switches are attractive for applications in data-centers and in high-performance computers due to their promise for high-speed switching, small footprints and low-power consumptions\cite{Cheng2018-ct,Batten2009-ke, nikolova2017modular}. Here, by programming the rings of a $16\times16$ silicon ring resonator-based switch, we demonstrate the automatic programming of a large system where the rings are inter connected via bus waveguides. 
Fig.~\ref{fig3}a shows a microscope picture of the fabricated switch. The $16\times16$ switch consists $256$ unit cells arranged in a cross-grid\cite{Poon2009-mn,Batten2009-ke} packed into a compact area of $6.11$ mm$^2$. In comparison, an $8\times8$ switch based on Mach-Zehnder interferometers occupied $8.25$ mm$^2$ in a prior-art\cite{Annoni2016-qb}. Our design can be made further compact by shrinking the contact pads, here designed to be \SI{80}{\micro\meter}$\times$\SI{80}{\micro\meter} for ease of prototyping. Fig.~\ref{fig3}b shows a microscope picture of a unit cell. {\ifHLcorrections{\color{red}\fi The \SI{8}{\micro\meter} radius ring resonator and waveguide crossing used in the unit cell is similar to that illustrated in Fig.~\ref{fig2}.}

\begin{figure}[h!]
	\centering
	\includegraphics[width=\textwidth]{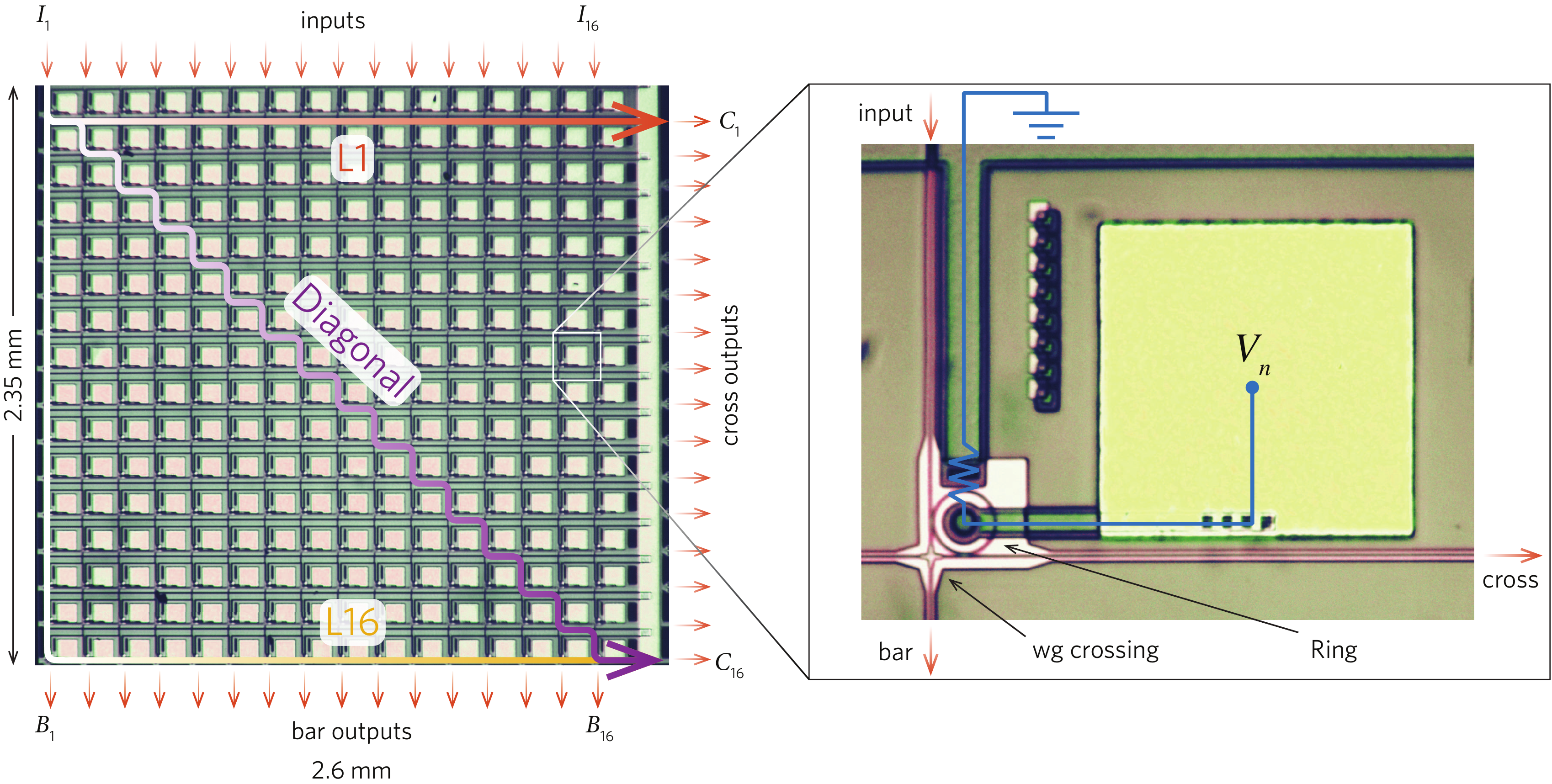} 
	\caption{{$16\times16$ ring resonator switch.}  Microscope picture of the fabricated switch showing the inputs-($I_1$-$I_{16}$), and bar-($B1-B_{16}$) and cross-($C_1-C_{16}$) outputs. The three routing configurations demonstrated in this paper are indicated as diagonal, $L_1$, and $L_{16}$. The inset shows the microscope picture of a unit cell. The location of photoconductive heater is shown as a resistor in the circuit.}
	\label{fig3}
\end{figure}

In order to demonstrate that we can automatically route any input-port to any cross/bar-port, we chose to automatically align the resonances of the $31$ rings along the longest route of the switch (labeled as diagonal in Fig.~\ref{fig3}) from input-port $I_1$ to cross-port $C_{16}$. This was accomplished by sequentially tuning the rings along the diagonal, starting from the ring nearest to the input-port to maximize the photocurrent in each ring.
Before tuning, the dark current or the heater current, $I_\text{heater}$, was measured and recorded for all the heaters in the voltage range of interest. As described previously, during the tuning process, the monitor photocurrent $I_\text{PD}$ of each ring was measured by subtracting the calibrated $I_\text{heater}$ from the total current measured. 
Limited by our measurement setup (see \textcolor{urlblue}{Supplementary section 3}), we only tuned one ring at a time and the voltages of the previously tuned rings were held constant. 
When rings $8, 16, 24, 31$ along the the diagonal path were reached, we repeated the tuning starting from ring $1$ to adjust for any detunings due to thermal and electrical crosstalk.
%
%
In an industrial application, this re-tuning can be avoided by continuously seeking the maximum photocurrents in the rings simultaneously{\ifHLcorrections{\color{red}\sout{by employing a maximum search algorithm}}\fi. It is important to note that in our system, the feedback loop controlling each ring is local to itself. 
{\ifHLcorrections{\color{red}\fi Therefore, when tuning all of the rings simultaneously, each ring's control circuit would detect and correct for any deviation of its resonance from the channel wavelength, including any perturbations due to thermal crosstalk from surrounding heat sources}.
Hence, this approach can be readily scaled to systems with any number of rings, as long as there is {\ifHLcorrections{\color{red}\fi sufficient light inside the rings for accurately sensing their resonance conditions}.

\begin{figure}[b]
	\centering
	\includegraphics[width=\textwidth]{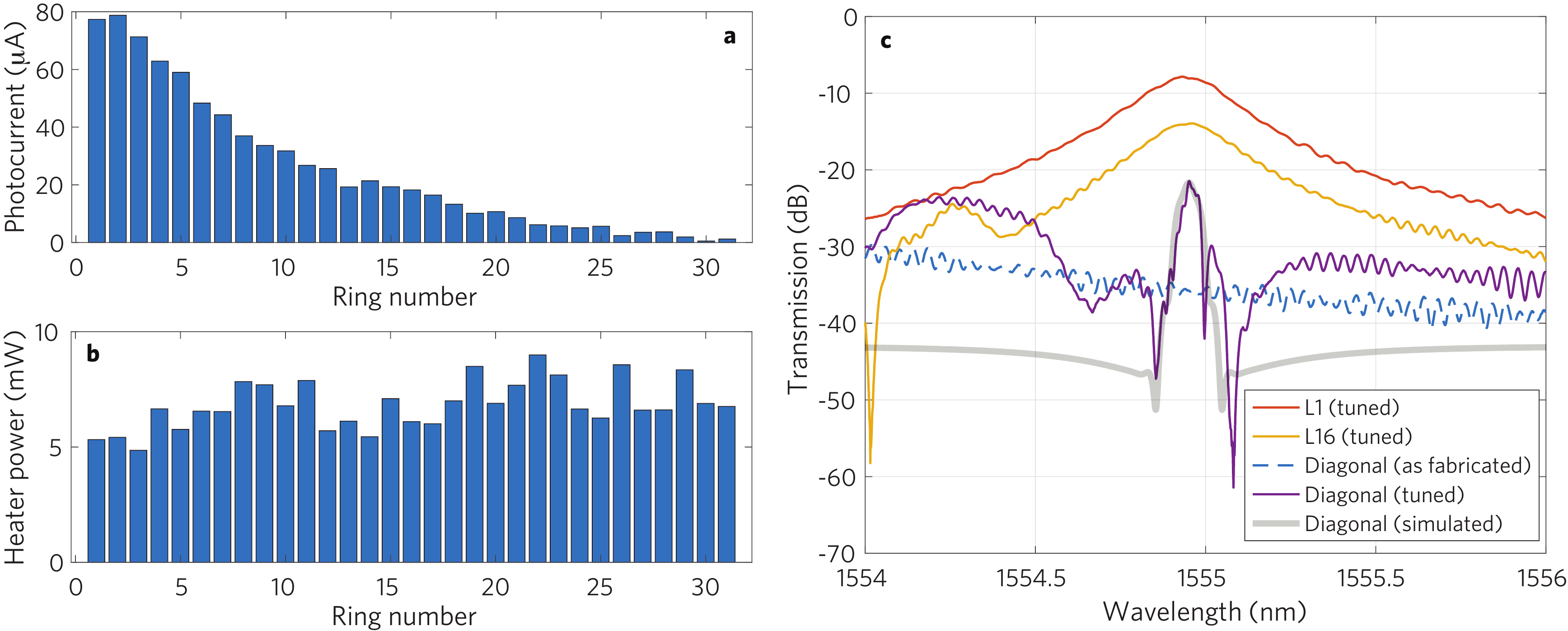} 
	\caption{{Programming the switch. (a)} Maximum photocurrent measured in each ring's photoconductive heater when tuning the $31$-rings along the diagonal from $I_1$-$C_{16}$. (b) shows the corresponding electrical powers supplied to the heaters which maximized the photocurrents. (c) Measured spectral responses for the as-fabricated and the after-configured switch along $L_{1}$, $L_{16}$, and diagonal paths. The simulated transfer function for the diagonal agrees well with the measured response near the peak wavelength. However, a mismatch exists away from the peak wavelength. This is because in simulation we assumed that all of the rings not in the diagonal were tuned away from the peak wavelength by half of the FSR (i.e., turned-off). However, in the experiment, the resonances of these rings were not controlled and the parasitic pathways formed through these rings increased the optical power collected away from the peak wavelength.}
	\label{fig4}
\end{figure}

Figs.~\ref{fig4}a and \ref{fig4}b shows the maximum photocurrents measured in each ring and the heater powers required for tuning the 31 rings of the diagonal route {\ifHLcorrections{\color{red}\fi from port $I_1$ to $C_{16}$}. The total power consumption of the rings' after tuning was $212$ mW. Fig.~\ref{fig4}c shows the measured spectral response from port $I_1$ to $C_{16}$ as fabricated (blue) and after tuning the {\ifHLcorrections{\color{red}\fi diagonal route} (purple). Fig.~\ref{fig4}c also shows the spectral responses measured by routing light via routes $L_1$ and $L_{16}$. Configuring paths $L_{1}$ and $L_{16}$ only required tuning a single resonator to be resonant with the channel wavelength.  Using these measurements, we estimated the insertion loss of a single ring and a crossing to be $0.3$ dB and $0.4$ dB, respectively (see \textcolor{urlblue}{Supplementary Fig. S5}).
The diagonal route, which routed light through $30$ crossings and $31$ rings, suffered the highest insertion loss.
In future, these losses can be reduced by design improvements\cite{Celo2017-hz}. In fact, the high insertion loss further indicates the capability of our photoconductive heaters in accurately measuring optical powers across a  dynamic range exceeding $21.3$ dB, where the input powers to the first and last rings in the path were estimated to be $7.2$ dBm and  $-14.1$ dBm, respectively. As shown in Fig.~\ref{fig4}c, the simulated transmission spectra of the diagonal including the insertion loss of the rings and crossings agree well near the peak wavelength with the measured spectral response.
The mismatch away from the peak wavelength is caused by the the parasitic pathways formed through untuned rings of the system which increased the optical power collected away from the peak wavelength in the experiment.
{\ifHLcorrections{\color{red}\fi Light in such parasitic pathways can be minimized by tuning the rings not belonging to the light path away (`turned-off') from their resonances. The simulated result for the Diagonal path in Fig.~\ref{fig4}c shows this case, where we have assumed that all of the rings not in the diagonal were tuned away from the peak wavelength by half of the FSR (i.e., turned-off)}.

\section{Tuning of 14-ring CROW}

Coupling ring resonators to form CROWs is an attractive approach for designing devices such as optical filters with flat pass-bands and steep roll-offs, optical delay lines, and four-wave mixing elements for a range of applications from sensing to quantum computing\cite{Morichetti2012-hz,Kumar2014-yi,Wang2014-tl}. 
Due to the coupling between their rings, resonance conditions of the rings of a CROW cannot be readily determined from its outputs alone. In this work, by embedding photoconductive heaters into each of the rings of a CROW,  we directly probe the rings' intracavity powers to find the desired resonance conditions using a simple tuning technique.
We demonstrate automatic resonance alignment of a $14$-ring CROW, correcting for the unwanted resonance shifts of its rings due to fabrication variations. 
Fig.~\ref{fig5}a shows a microscope picture of the fabricated $14$-ring CROW. 
The inset illustrates the integration of photoconductive heaters into each of the rings. The rings, starting from the ring coupled to the input-port, are denoted as R$_1$ through R$_{14}$. The ring-ring and ring-bus waveguide couplings were chosen for a maximally flat drop-port response (see \textcolor{urlblue} {Supplementary section 1}). The area occupied by the 14 rings and their photoconductive heaters is only $13.5\times10^{3}$ \SI{}{\micro\meter\squared}. 

\begin{figure}[b]
	\centering
	\includegraphics[width=\textwidth]{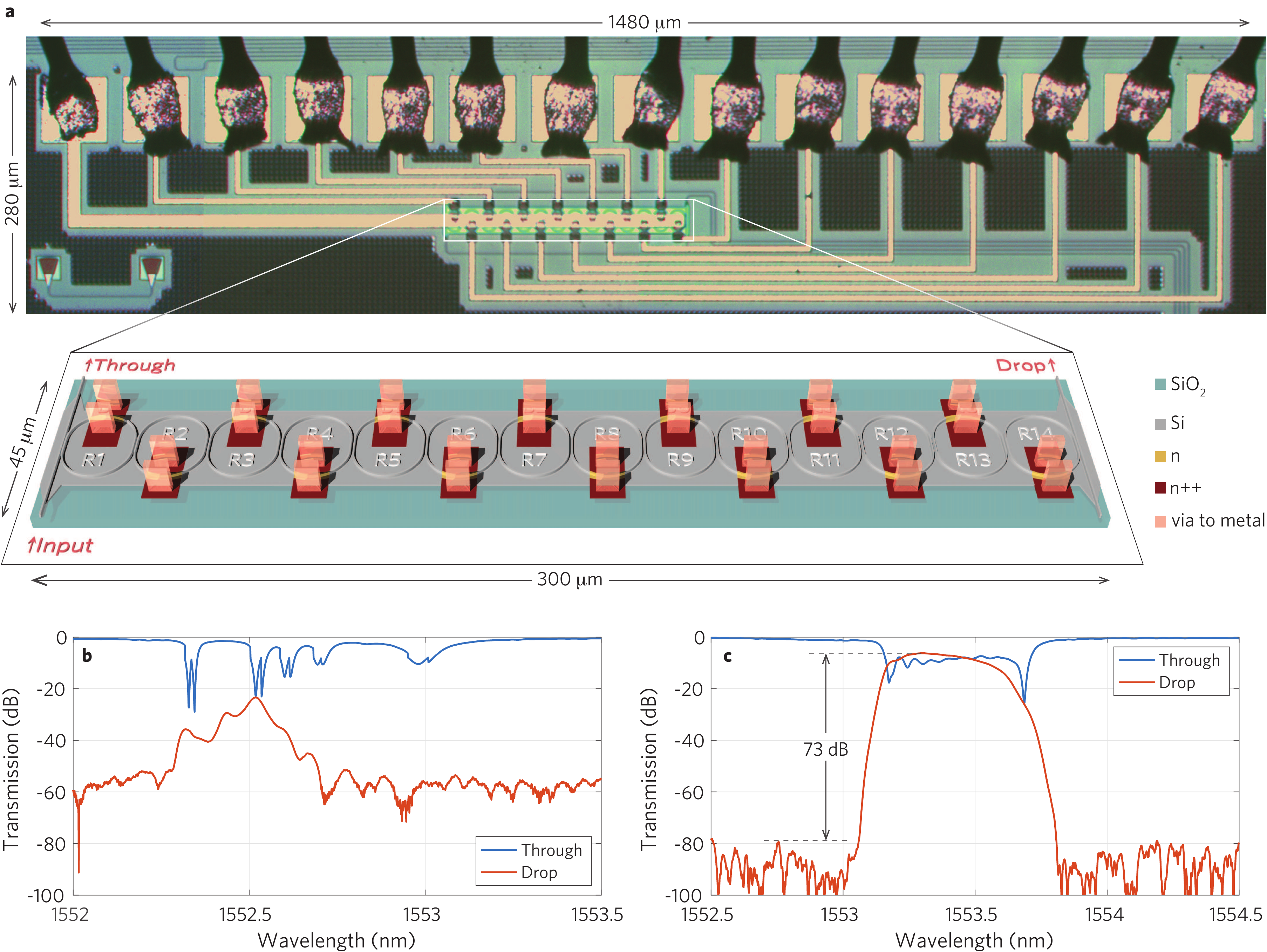} 
	\caption{{Resonance alignment of $14$-ring CROW. (a)} Microscope picture of the fabricated $14$-ring CROW. The inset illustrates the layout of photoconductive heaters in the rings. {(b)} Measured as-fabricated through- and drop-port spectra of the CROW. {(c)} Measured through- and drop-port spectra after tuning.}
	\label{fig5}
\end{figure}

In order to automatically tune the CROW to the input laser's wavelength, first, we tuned rings R$_1$ through R$_{14}$ to maximize the photocurrent in each ring's photoconductive heater. This step brought the resonances of all the rings closer to the input laser's wavelength, which ensured that sufficient photocurrent can be detected in each ring for the subsequent tuning steps. 
Next, we tuned the rings in reverse order from R$_{14}$ through R$_1$. 
During each tuning step, while tuning the $n^\text{th}$ ring (R$_n$), we measured photocurrents of rings R$_{n-1}$ and R$_n$. 
The heater power corresponding to the desired resonance condition of R$_n$ was found by maximizing the ratio of the photocurrents $I_{\text{PD},n}/I_{\text{PD},n-1}$, where $I_{\text{PD},n}$ is the photocurrent measured in the $n$\textsuperscript{th} ring\cite{Jayatilleka2018-ed}. When R$_1$ was reached, we maximized $I_{\text{PD},1}$ to find its resonance condition. (\textcolor{urlblue}{Supplementary Fig. S6} shows the measured photocurrents and their ratios $I_{\text{PD},n}/I_{\text{PD},n-1}$ for these $14$ tuning steps). Finally, we iterated over these 14 tuning steps several times in order to adjust for any undesired detunings due to the thermal crosstalk between the rings. 
{\ifHLcorrections{\color{red}\fi In simulation, we verified that the desired solution can be reached by iterating the tuning steps in the above described sequence. For example, detuning due to ambient temperature changes can be corrected by continuously cycling through the tuning steps.}
The feedback loops used here for finding the resonance conditions of the rings are local to the adjacent rings of the CROW, {\ifHLcorrections{\color{red}\fi  i.e., the desired resonance condition of a ring  can be found only by sensing a ring and its neighbor without knowing/sensing the resonance conditions of the other rings of the CROW.  \ifHLcorrections{\sout{which makes this method applicable for tuning CROWs regardless of the number of rings they contain}}\fi Hence, this localization paves the path towards tuning CROWs with even higher number of rings by minimizing the requirement of a `supervisory' control unit.}

Fig.~\ref{fig5}b shows the as-fabricated through- and drop-port responses of the CROW and Fig.~\ref{fig5}c shows the improved responses after tuning. While the theoretically simulated extinction ratio of this device exceeds $200$ dB, the measurable extinction ratio here is about $73$ dB due to limitations with the measurement instruments and the insertion loss of the grating couplers. 
The noise floor of the post tuning spectra was reduced by averaging over $1000$ consecutive spectral sweeps collected using the optical vector network analyzer\cite{Ong2013-zw}.
After tuning, the $14$-rings consumed a total power of $64.9$ mW. Insertion loss of the tuned CROW at the center of the drop-port pass-band was $7.15$ dB. {\ifHLcorrections{\color{red}\fi The FSR of the tuned filter was {8.4}{ nm} at 1553.4 nm wavelength.}

\section{Conclusion}

We showed record high photoconductive QEs (up to $200$) in doped silicon waveguides. Using such photoconductive heaters, we demonstrated the automatic alignment of 31 ring resonators along the longest path of a $16\times16$ switch and the tuning of a 14-ring CROW. These are the largest and, yet, most compact automatically tuned silicon ring resonator circuits to date. The high QE and the large dynamic range of the  photoconductive heaters allowed the resonance conditions of individual resonators of these systems to be precisely sensed and tuned simultaneously without the need for additional material depositions (zero change to foundry fabrication process\cite{noauthor_undated-re}) or photodetectors, complex tuning algorithms, and without increasing the number of contact pads.
As a result of this increased insight into the resonance conditions of the individual resonators of the system, the tuning methods we used localized the feedback loops to individual resonators of the switch and to the adjacent resonators of the CROW. 
{\ifHLcorrections{\color{red}\fi Therefore, these methods are readily applicable for tuning a ring resonator-based system regardless of the number of resonators,  as long as there is sufficient light in each resonator for sensing the resonance conditions, i.e., $>-35$~dBm for the work presented here.}
By providing a highly scalable and a low-cost solution which preserves the miniature sizes of silicon ring resonators, our results indicate a path forward for making a multitude of long-promised ring resonator systems demonstrated in previous works over the past two decades viable in practice \cite{Batten2009-ke,Atabaki2018-dv,Khilo2012-rk,Wang2014-tl,Kumar2014-yi,Tait2017-so,Morichetti2012-hz, nikolova2017modular, khope2017chip}.

\section*{Acknowledgement}
The authors thank Wim Bogaerts, Nicolas A. F. Jaeger, Andrea Melloni, Antonio Ribeiro, Mustafa Hammood, and Anthony Park for their help, Stanley Shang of Testforce Systems for lending an Optical Vector Analyzer, CMC microsystems for providing access to fabrication and design tools, and the Natural Sciences and Engineering Research Council of Canada, the SiEPIC program,  and CMC microsystems for financial support.

\bibliographystyle{osajnl}
\bibliography{references}

\end{document}